\documentclass[amssymb,aps,amsmath,floatfix,pre,twocolumn,superscriptaddress,longbibliography]{revtex4-2}
\usepackage[latin1]{inputenc}

\usepackage[T1]{fontenc}
\usepackage{graphicx}
\usepackage{amsmath,amssymb,amsfonts, MnSymbol}
\usepackage{mathrsfs}
\usepackage{bm}
\usepackage{hyperref}
\usepackage{bbold}
\usepackage{xcolor}
\usepackage{tikz}

\newcommand{\algn}[1]{\begin{align} #1 \end{align}}

\newcommand{\eps}{\ensuremath{\varepsilon}}

\newcommand{\ve}[1]{\boldsymbol{#1}}
\newcommand{\nn}{\nonumber}
\newcommand{\ee}{\ensuremath{\text{e}}}
\newcommand{\ed}{\ensuremath{\text{d}}}

\newcommand{\dd}[1]{\ensuremath{\tfrac{\text{d}}{\text{d} #1}}}

\newcommand{\VddVT}{V_{\text{dd}}/V_\text{T}}

\newcommand{\VddcVT}{V^*_{\text{dd}}/V_\text{T}}

\newcommand{\Vdd}{V_{\text{dd}}}
\newcommand{\Vddc}{V^*_{\text{dd}}}

\newcommand{\1}{\textbf{1}}

\newcommand{\mc}[1]{\ensuremath{\ms{#1}}}
\newcommand{\ms}[1]{\ensuremath{\mathscr{#1}}}
\newcommand{\mbb}[1]{\ensuremath{\mathbb{#1}}}
\newcommand{\kB}{\ensuremath{k_\text{B}}}
\newcommand{\eqnlab}[1]{\label{eq:#1}}
\newcommand{\seclab}[1]{\label{sec:#1}}

\newcommand{\eqnref}[1]{\eqref{eq:#1}}
\newcommand{\Eqnref}[1]{Eq.~\eqref{eq:#1}}
\newcommand{\Eqsref}[1]{Eqs.~\eqref{eq:#1}}
\newcommand{\secref}[1]{\ref{sec:#1}}
\newcommand{\Secref}[1]{Sec.~\ref{sec:#1}}
\newcommand{\figref}[1]{\ref{fig:#1}}
\newcommand{\Figref}[1]{Fig.~\ref{fig:#1}}
\newcommand{\Figsref}[1]{Figs.~\ref{fig:#1}}

\begin{document}
\title{Dissipation and fluctuations of CMOS ring oscillators close to criticality}

\author{Ashwin Gopal}
\affiliation{Complex Systems and Statistical Mechanics, Department of Physics and Materials Science, University of Luxembourg, 30 Avenue des Hauts-Fourneaux, L-4362 Esch-sur-Alzette, Luxembourg}
\author{Massimiliano Esposito}
\affiliation{Complex Systems and Statistical Mechanics, Department of Physics and Materials Science, University of Luxembourg, 30 Avenue des Hauts-Fourneaux, L-4362 Esch-sur-Alzette, Luxembourg}
\author{Jan Meibohm}
\affiliation{Theoretical Physics Unit, Institute for Physics and Astronomy, Technische Universit\"at Berlin, Hardenbergstra\ss{}e 36, 10623 Berlin, Germany}

\begin{abstract}
We analyze a thermodynamically consistent model of CMOS-based ring oscillators near the onset of coherent voltage oscillations. For driving voltages close to the critical value, we derive the normal form of the Hopf bifurcation that underlies the oscillation transition in the thermodynamic limit. Using this normal form, we determine the phase and amplitude dynamics, and demonstrate that entropy dissipation decreases in the oscillating state for ring oscillators with more than three inverters. These findings culminate in a stability-dissipation relation, which links the observed decrease in dissipation to an increase in the local stability of the oscillating state. Next, we characterize finite-size fluctuations of the amplitude and phase near the critical voltage, using a stochastic version of the normal form. We demonstrate that close to the transition, finite-size fluctuations remain important at arbitrary system size, introducing oscillations even below the critical voltage. We further show that these noise-induced oscillations have an anomalously short decoherence time that scales sub-linearly with the system-size, in contrast to the behavior far from criticality.
\end{abstract}
\maketitle
\section{Introduction}
Self-sustained oscillations are paradigmatic phenomena in non-equilibrium physics, manifested across systems ranging from biochemical~\cite{goldbeter1996biochemical,novak2008design}, and electronic clocks~\cite{strogatz2024nonlinear,milburn2020thermodynamics}, to nonlinear optics~\cite{agrawal2000nonlinear,buks2019self} and hydrodynamics~\cite{rockwell1978self}. Driven far from equilibrium, these systems undergo a qualitative change from stationary states, whose probability distributions are concentrated near stable fixed points, to oscillating states, concentrated around the orbits of stable limit cycles. In the thermodynamic limit, this dynamical transition is sharp and well-described by deterministic bifurcation theory ~\cite{golubitsky1988singularities,nicolis1995introduction,strogatz2024nonlinear}. Recent work has revealed relationships between dynamical stability and dissipation, close to criticality at the deterministic level~\cite{meibohm2024minimum,meibohm2024small}.
However, on the mesoscopic scale, where the magnitude of thermal fluctuations is comparable to the size of the order parameter, the distinction between deterministic dynamics and thermal noise blurs. At this ``edge of instability,'' emerges a regime of rich physical interplay, with complex dynamical and thermodynamical behaviors~\cite{hohenberg1977theory,epstein1996nonlinear,herpich2018collective,nguyen2018phase, gopal2024information}.

Complementary metal-oxide-semiconductor (CMOS) circuits~\cite{taur2021fundamentals} provide a practical platform to investigate this interplay between deterministic and stochastic dynamics, and its thermodynamic implications~\cite{freitas2021stochastic,gao2021principles,gopal2024nonlinear,helms2025stochastic}. As CMOS transistors scale down, with smaller physical dimensions and lower supply voltage, various sources of physical noise become relevant to the signal. This is particularly acute in ultra-low-power applications~\cite{schrom1996ultra, kim2011ultra} and neuromorphic computing~\cite{mead2002neuromorphic,milo2020memristive,jelinvcivc2025efficient}, where circuits often operate in the subthreshold regime with supply voltages approaching the thermal voltage. 
Among the diverse digital, analog, and mixed-signal components built from CMOS transistors, ring oscillators constitute a fundamental building block for timing, frequency generation, and signal processing. Their minimal design, high tunability, and compatibility with standard fabrication processes make them indispensable in both high-performance and ultra-low-power applications~\cite{baker2019cmos}.

\begin{figure}
\centering
\includegraphics[width=\linewidth]{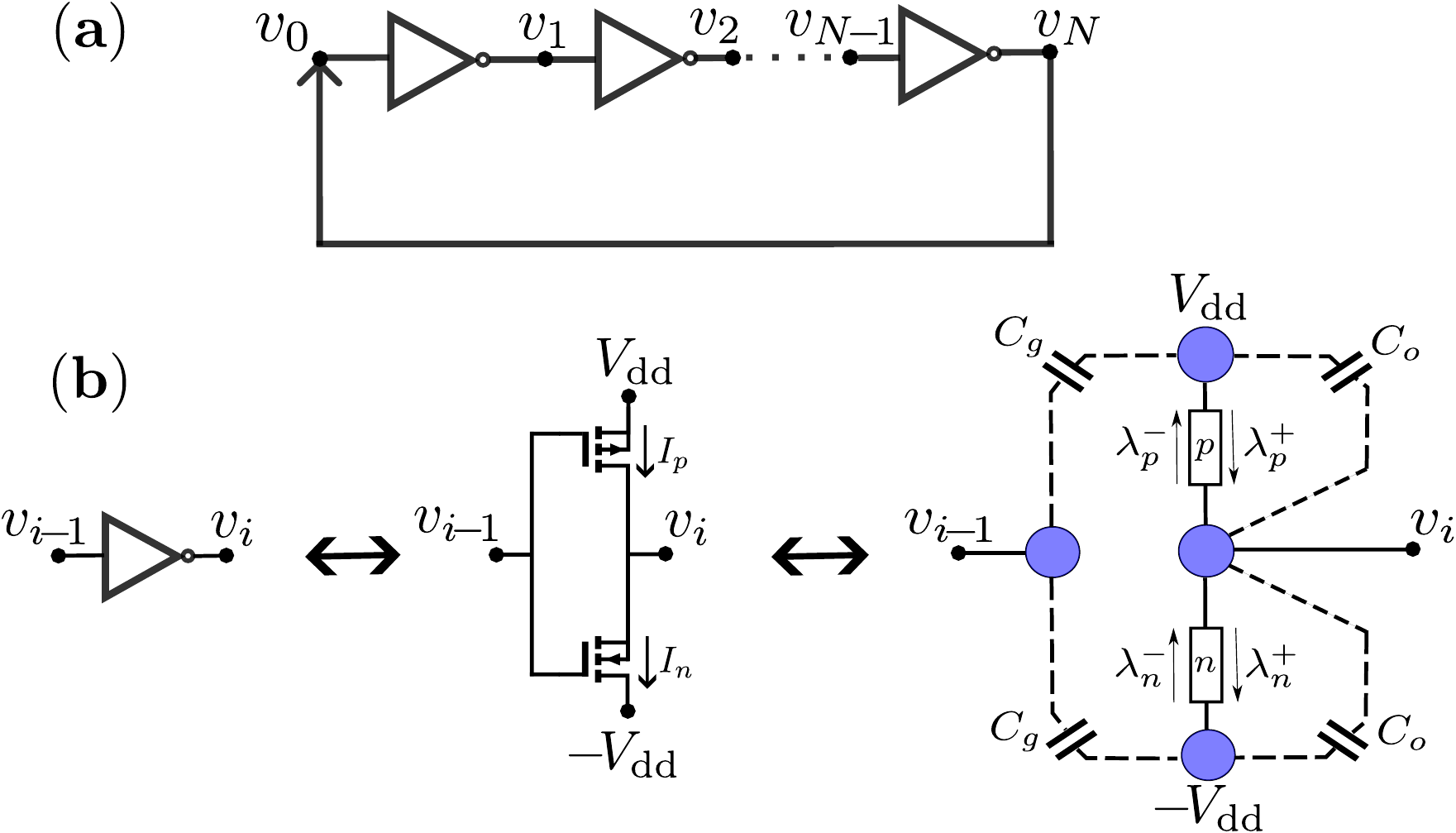}
\caption{(a) Logical representation of a ring oscillator as an odd number $N$ of inverters (NOT gates) connected in a loop, i.e $v_N=v_0$. (b) CMOS implementation of the NOT gate (inverter) using pMOS and nMOS transistors, and a bi-Poissonian charge-transport model including the gate-body $C_g$ and drain-source $C_o$ capacitances~\cite{freitas2021stochastic}.}
\label{fig:ring}
\end{figure}

Ring oscillators are electronic devices composed of an odd number of inverters connected in a loop~\cite{razavi2020design,mandal2010ring}; see \Figref{ring}. Their output voltages oscillate because of an inherent instability in the circuit's logical states. This instability drives the loop into self-sustained oscillations, whose frequency is determined by the number of inverters and the propagation delay between them. This delay, in turn, is controlled by the voltage across the inverters, so ring oscillators are highly tunable and therefore commonly serve as voltage-controlled oscillators in phase-locked loops~\cite{razavi2020design,mandal2010ring}. Their tunability and simplicity make ring oscillators essential components in CMOS technology, particularly in timing and frequency-synthesis applications.

Current fluctuations in the transistors -- thermal or flicker -- induce random variations of the oscillation phase producing phase noise, and causing ring oscillators to lose coherence over time, manifested as jitter~\cite{hajimiri2002jitter}. Although detrimental in clocking contexts, phase noise can be exploited to design ``true random number generators''~\cite{robson2014truly,bucci2003high,sunar2006provably}. Unlike pseudo-random number generators, these devices rely on inherently noisy physical processes and are used in secure communication and cryptographic systems where unpredictability is essential~\cite{liu2016low}. True randomness can be generated by combining the uncertainties of multiple ring oscillators to produce a bit stream at specified sampling intervals. In this way, amplified phase noise is harnessed to construct efficient true random number generators even in the subthreshold regime of ultra-low power~\cite{guler2012modeling,guler2013modeling}.

In this paper, we analyze a thermodynamically consistent model of ring oscillators near the critical voltage, at which small-amplitude oscillations emerge. We derive the normal form for the Hopf bifurcation of the dominant oscillating mode that  underlies the transition, enabling us to determine the amplitude and frequency of oscillations close to onset. Using the framework of stochastic thermodynamics~\cite{seifert2012stochastic,van2015ensemble,pelitiStochasticThermodynamicsIntroduction2021}, we show that entropy production decreases in the oscillatory state for ring oscillators with more than three inverters. Based on these findings, we derive a stability-dissipation relation, similar to that in Refs.~\cite{meibohm2024minimum,meibohm2024small}, linking the reduced entropy production to the enhanced local stability of the oscillating state.

We then characterize phase noise near the critical voltage, where conventional large-amplitude approaches~\cite{demir1998phase,vance1996fluctuations,gaspard2002correlation} become inaccurate. Our analysis is based on a ``stochastic normal form''~\cite{xiao2007effects} that incorporates the leading noise-induced effects. Close to the oscillation transition, such effects remain relevant, regardless of the large system size. In particular, they generate noise-induced oscillations even below the critical voltage. In contrast to large-amplitude oscillations~\cite{gaspard2002correlation,remlein2022coherence}, these oscillations exhibit unusually strong decoherence with an anomalously short coherence time. This property may indicate that noise-induced oscillations could be especially well suited for generating high-rate, statistically independent true random numbers.

The paper is organized as follows. In \Secref{model} we introduce the thermodynamically consistent formulation of electric circuits of Ref.~\cite{freitas2021stochastic} and describe our model. In \Secref{detlim} we analyze the thermodynamic limit, identify the equilibrium fixed point and its instability at the critical voltage, and derive the Hopf normal form that characterizes small-amplitude oscillations. We then discuss the stability and thermodynamics of the deterministic system. Section~\secref{deviations} addresses finite-size fluctuations: Here, we analyze the stochastic normal form, noise-induced oscillations, and the resulting (short) decoherence time. Finally, we present our conclusions in~\Secref{conc}.
\section{Model}\seclab{model}
We consider a CMOS-based ring oscillator composed of (odd) $N$ inverters [see Fig.~\ref{fig:ring}(a)], in contact with a heat bath at temperature $T$. Each inverter is composed of two metal-oxide-semiconductor field-effect transistors (MOSFETs) [Fig.~\ref{fig:ring}(b), center], one $n$-doped (``nMOS'', bottom) and one $p$-doped (``pMOS'', top). The heat bath induces finite-temperature voltage fluctuations in the system. In order to consistently describe them, we need to account for the voltage fluctuations that arise from all $2N$ MOSFETs in the circuit.

Using the formalism in Ref.~\cite{freitas2021stochastic}, we model MOSFETs as conduction channels subject to thermal noise. We consider the subthreshold regime in which voltage changes across the transistors are caused by single-electron transitions through the conduction channels. Hence, electron transitions through the nMOS and pMOS transistors of the $i$th inverter are treated as bidirectional Poisson processes, characterized by a forward ($+$) and a backward ($-$) transition rate $\lambda^{\text{n/p}}_{i\pm}(v_{i-1}, v_{i})$. Here, $v_i$ denotes the output voltage of the $i$th inverter.

The inverters are connected in a loop, see Fig.~\ref{fig:ring}(a), so the input of the first inverter is the output of the last inverter, i.e. $v_{N}=v_0$ and, more generally, $v_i =v_{i\!\!\!\mod N}$. Because MOSFETs typically have insulating gate terminals, changes in the $i$th component $v_i$ of the output voltage are only due to charge transitions in the transistors of the $i$th inverter.

The allowed transitions in the voltage vector $\ve v = (v_0,\ldots,v_{N-1})^{\sf T}$ are given by
\algn{\eqnlab{alltransN}
	\ve v\to \ve v+\ve \Delta^{\text{p}/\text{n}}_{i\pm} v_e\,,
}
with
\algn{
	\quad (\ve \Delta^\text{p}_{i\pm})^m = \pm\delta_{im}\quad \text{and}\quad  (\ve \Delta^\text{n}_{i\pm})^m = \mp\delta_{im}\,,
}
where $\delta_{mn}$ is the Kronecker delta and $v_e \equiv q_e/C$ represents the elementary change of the output voltage due to a single charge transfer. Furthermore, $C= 2(C_o + C_g)$ describes the effective capacitance at the output node with contributions from the gate-body capacitance $C_g$ and additional output capacitance $ C_o$ caused by the transistors connected to the node.

The current $I_{p}$ through a pMOS transistor as a function of the input $V_{\text{in}}$, output $V_{\text{out}}$ and driving voltages $\Vdd$ is given by~\cite{freitas2021stochastic}
\begin{multline}
    I_{p}(V_{\text{in}}, V_{\text{out}};V_{\text{dd}})=\\
   I_0\ee^{(V_{\text{dd}}-V_{\text{in}}-V_{\text{th}})/V_\text{T}}\left[1-\ee^{-(V_{\text{dd}}-V_{\text{out}})/V_\text{T}}\right]\,, \eqnlab{Ip}
\end{multline}
where $V_{\text{th}}$ denotes the threshold voltage and $I_0$ is the characteristic current, which are parameters determined by the transistor specifications.

We consider CMOS devices with symmetric transistor parameters for pMOS and nMOS transistors and further apply the symmetric powering voltages $V_{\text{dd}}$ and $-V_{\text{dd}}$, respectively. In such a setup, the current $I_{n}$ through an nMOS transistor is obtained by inverting the input and output voltages in \Eqnref{Ip}, i.e.,
\begin{equation}
	I_{n}(V_{\text{in}}, V_{\text{out}};V_{\text{dd}})= I_{p}(-V_{\text{in}}, -V_{\text{out}};V_{\text{dd}})\,.\eqnlab{In}
\end{equation}
The deterministic $I$-$V$ characteristics \eqnref{Ip} and \eqnref{In} together with the local detailed balance condition, required for thermodynamic consistency according to stochastic thermodynamics~\cite{seifert2012stochastic,van2015ensemble,pelitiStochasticThermodynamicsIntroduction2021}, uniquely determine the transition rates $\lambda^{\text{n/p}}_{i\pm}(v_{i-1}, v_{i})$ for all transistors~\cite{freitas2021stochastic}. The transition rates for the $i$th pMOS transistor of the ring oscillator can be written as
\algn{\eqnlab{jump_rates}
\lambda^\text{p}_{i+} (\ve v) &= \frac{\Omega}{\tau_0 }   \ee^{(V_\text{dd} -
v_{i-1})/V_{\text{T}}}\nonumber\,,\\
\lambda^\text{p}_{i-}(\ve v) &= \frac{\Omega}{\tau_0 } \ee^{(v_{i}-v_{i-1})/V_\text{T}}\ee^{-1/2\Omega}\,,
}
where we have defined the system size parameter $\Omega=V_\text{T}/v_e=k_B T C/q_e^2$ and the characteristic timescale $\tau_0 = [(V_\text{T} C)/I_0]\ee^{V_{\text{th}}/V_\text{T}}$ with the thermal voltage $V_\text{T} =k_B T/q_e$. The previously described symmetries between the MOSFETs imply that the transition rates for nMOS are given by
\begin{equation}
	\lambda^\text{n}_{i\pm}(\ve v) = \lambda^\text{p}_{i\pm}(-\ve v)\,.
\end{equation}

Combining these ingredients, the dynamics of the probability $P(\ve v,t)$ to find the system in state $\ve v$ at time $t$ is given by the Master equation
\begin{multline}\eqnlab{mastereqnnonint}
	\partial_t P(\ve v,t) 	= \sum_{i=0}^{N-1} \sum_{\rho =n,p} \left\{\lambda_{i+}^{\rho }(\ve v - v_e \ve{\Delta}_{i+}^{\rho })P(\ve v - v_e \ve{\Delta}_{i+}^{\rho }, t)\right.\\
	\left.+\lambda_{i-}^{\rho }(\ve v - v_e \ve{\Delta}_{i-}^{\rho })P(\ve v - v_e \ve{\Delta}_{i-}^{\rho }, t)\right.\\
	\left.- \left[\lambda_{i+}^{\rho }(\ve v) + \lambda_{i-}^{\rho }(\ve v)\right]P(\ve{v}, t)\right\}\,.
\end{multline}
\begin{figure}
\centering
\includegraphics[width=\linewidth]{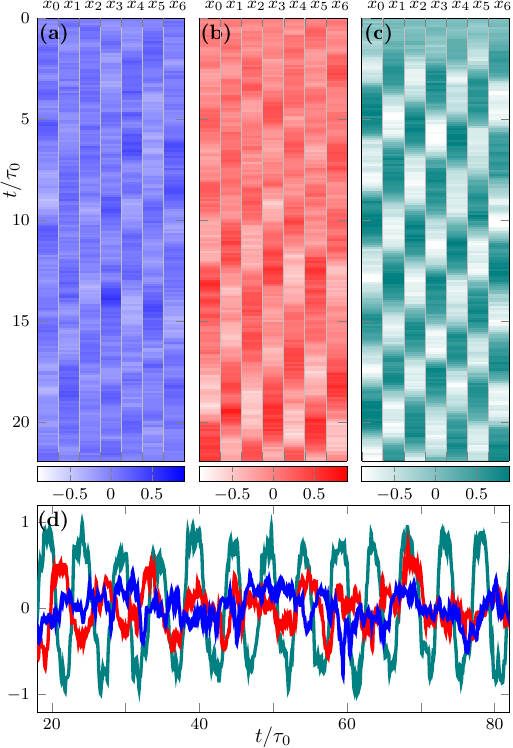}
\caption{Dimensionless output voltage vector $\ve x = \ve v/V_\text{T}$ as function of time $t$ below and above the critical voltage $\VddcVT \approx 0.7466$ for a $7$-stage ring oscillator with system size $\Omega=10^2$. \textbf{(a)} $\ve x$ as function of $t$ for $\lambda = (V_\text{dd}-\Vddc)/V_\text{T} = -0.1$ as color map. Darker colors correspond to higher voltages \textbf{(b)} $\lambda = 0$. \textbf{(c)} $\lambda = 0.2$. \textbf{(d)} Voltage $v_0$ as function of $t$ for $\lambda=-0.1$ (blue), $\lambda = 0$ (red), and $\lambda=0.2$ (green).
}
\label{fig:sample_trajecs}
\end{figure}
Figure~\ref{fig:sample_trajecs} shows sample trajectories for the output voltage $v_0$ of a 7-stage ring oscillator for different driving voltages $\Vdd$. We observe that for small $\Vdd$ the stochastic trajectories fluctuate incoherently around $\ve v = 0$, with small amplitudes (blue and red in \Figref{sample_trajecs}). Upon increasing $\Vdd$ across a critical voltage $\Vddc$ to be determined, the oscillations become coherent and grow in amplitude (green in \Figref{sample_trajecs}). We explain this behavior in detail in the next Sections.
\section{Small amplitude oscillations in thermodynamic limit}\seclab{detlim}
The onset of coherent oscillations upon increasing $\Vdd$, observed in \Figref{sample_trajecs}, is most conveniently described in the thermodynamic limit $\Omega \to \infty$ of large system size. Physically, this limit can be realized by simply increasing the physical dimensions of the transistor~\cite{gopal2022large}, so that the elementary voltage change $v_e$  becomes negligible compared with all other voltages. Since the current flow scales with the physical dimensions of the channel, we define rescaled transition rates $\omega^{\pm}_{\text{n/p}}(\ve v)$ as 
\begin{equation}
	\omega_{i\pm}^{\text{n/p}}(\ve v) \equiv \lim_{\Omega\to\infty}\frac{\lambda_{i\pm}^{\text{n/p}}(\ve v)}{\Omega}\,.
\end{equation}
To determine the dynamics in this limit, we expand the Master equation~\eqnref{mastereqnnonint} to lowest order in $\Omega^{-1}$ to obtain
\begin{equation} \label{eq:deterministic}
	\partial_tP(\ve v,t) =-\nabla_{\ve v}\cdot[\ve F(\ve v)P(\ve v,t)] +\mathcal{O}(1/\Omega)\,,
\end{equation}
where we have identified the deterministic drift
\begin{align}
	&\ve F(\ve v)	=\sum_{i=0 }^{N-1}\sum_{\rho=\text{n},\text{p} }\left[\ve{\Delta}_{i+}^{\rho } \omega^{\rho}_{i+} (\ve v)+ \ve{\Delta}_{i-}^{\rho } \omega^{\rho}_{i-}(\ve v)\right]\,,\\
	&		= \frac{2}{\tau_0}\left[\sinh\left(\frac{v_{i-1}(t)-v_{i}(t)}{V_\text{T}}\right)-\ee^{V_{\text{dd}}/V_\text{T}}\sinh\left(\frac{v_{i-1}(t)}{V_\text{T}}\right) \right]\,.\nn
\end{align}
Due to the absence of a diffusion term in \Eqnref{deterministic}, the trajectories of the output voltages $v_i$ follow a deterministic dynamics given by
\begin{equation}
     \frac{\dot v_i(t)}{V_\text{T}} = F_i(\ve v) = \tau_0^{-1}h\left(\frac{v_i}{V_\text{T}},\frac{v_{i-1}}{V_\text{T}}\right)\,,\eqnlab{eom}
\end{equation}
where $\dot v_i = \dd{t}v_i$ denotes a time derivative and the function
\algn{
	h(x,y) = 2\left[\sinh\left(y-x\right)-\ee^{V_{\text{dd}}/V_\text{T}}\sinh\left(y\right) \right] \,,
}
is identical for all $i$. In what follows, we discuss the basic properties of \Eqnref{eom}.
\subsection{Symmetry}
As one can anticipate by considering \Figref{ring}, ring oscillators composed of identical inverters are invariant under cyclic permutations of the inverters. This symmetry directly carries over to the output voltages $\ve v$, and is thus reflected in the dynamics~\eqnref{eom}.

For the ring oscillator, we define the symmetry operation $\gamma$ as $\gamma = \mathcal{R}^n$ for some integer $n \geq 0$, where the generator $\mathcal{R}$ cyclically permutes the components of $\ve{v}$ by one, i.e.,
\begin{equation}\label{permutation}
    (v_0, v_1, \ldots, v_{N-1})^{\mathrm{T}} \overset{\mathcal{R}}{\mapsto} (v_{N-1}, v_0, v_1, \ldots, v_{N-2})^{\mathrm{T}}.
\end{equation}
Symmetry under the action of $\gamma$ implies that the macroscopic dynamics of the system, described by \Eqnref{eom}, must be \textit{equivariant} with respect to the action of $\gamma$~\cite{golubitsky1988singularities}, which entails that
\begin{equation}\eqnlab{equivariance}
    \mathbf{F}(\ve{v}) = \gamma^{-1} \mathbf{F}(\gamma\, \ve{v}).
\end{equation}
This property can be seen to hold true by considering the right-hand side of \Eqnref{eom}, because the function $h$ is independent of the index $i$.

The cyclic permutation symmetry of inverters is a crucial property of the ring oscillator that we will use repeatedly in what follows.
\subsection{Trivial fixed point}\seclab{FP}
Since $\sinh(0) = 0$, one finds by inspection that \Eqnref{eom} admits a trivial fixed point when all voltages vanish, i.e., $\ve v^* = (0,0,\ldots,0)^{\sf T}$. Without external drive ($\Vdd=0$), the device is in thermal equilibrium, so that $\ve v^*$ must be stable. Driven devices ($\Vdd\neq0$), by contrast, are intrinsically out of equilibrium, so $\ve v^*$ may become unstable, giving rise to dynamic behavior. Generally, the stability of $\ve v^*$ is governed by the eigenvalues of the stability matrix $\mbb{M}$ with elements
\algn{\eqnlab{stab}
	M_{ij} = V_\text{T}\frac{\partial F_i(\ve v)}{\partial v_j}\bigg|_{\ve v = \ve v^*} = \tau_0^{-1}\left(h_{10}\delta_{ij} + h_{01}\delta_{i-1j}\right)\,,
}
where all indices are understood modulo $N$ and where we defined
\algn{
	h_{ij} = \partial_x^i\partial_y^jh(x,y)\big|_{x=y=0}\,,
}
so that
\algn{
	h_{10} = -2\,,\qquad h_{01} = -2(\ee^{\Vdd/V_\text{T}}-1)\,.
}
The fixed point $\ve v^*$ is considered stable when all eigenvalues of $\mbb{M}$ have a negative real part. If one or more of the eigenvalues have positive real parts, by contrast, perturbations along the corresponding eigendirections grow over time, and the fixed point is said to be unstable. To determine the conditions under which $\ve v^*$ is stable, we need to diagonalize the stability matrix $\mbb{M}$.

From \Eqnref{stab} we observe that $\mbb{M}$ is a circulant matrix~\cite{Dav79}, which means that $\mbb{M}$ commutes with $\gamma$,
\algn{\eqnlab{commutator}
	\gamma \mbb{M} =\mbb{M}\gamma\,.
}
This commutation relation, in turn, is a consequence of the equivariance property~\eqnref{equivariance} and thus of the symmetry of the ring oscillator. An important consequence of \Eqnref{commutator} is that the eigenvectors of \mbb{M} are discrete Fourier modes
\begin{equation}
    \hat{\ve v}_k = \left(1,\ee^{\frac{i2\pi k}{N}},\ee^{\frac{i4\pi k}{N}}\ldots,\ee^{\frac{i2\pi k(N-1)}{N}}\right)^{\sf T}
\end{equation}
for $k=0,1,\ldots,N-1$ and that $\mbb{M}$ is diagonalized by the discrete Fourier transform $\mbb{F}$, i.e.,
\algn{
	\mbb{D} = \mbb{FMF}^{-1},
}
where $\mbb{D}$ is the diagonal form of $\mbb{M}$ and $\mbb{F}$ and its inverse have the elements
\algn{\eqnlab{Ftrans}
	F_{kn} = \ee^{\frac{i2\pi kn}{N}}\,,\quad F^{-1}_{nk} = \frac1{N}\ee^{-\frac{i2\pi kn}{N}}\,.
}
The eigenvalues $\ms{D}_k = \mbb{D}_{kk}$ of $\mbb{M}$ are then obtained as $\mc{D}_k = \mu_k + i\omega_k$, where
\algn{
	\mu_k =& h_{10} + h_{01}\cos\left(\frac{2\pi k}{N}\right)\,,\nn\\
	=& -2\left[1+(\ee^{\VddVT}-1)\cos\left(\frac{2\pi k}{N}\right)\right]\,,
}
and
\algn{
	\omega_k &= h_{01}\sin\left(\frac{2\pi k}{N}\right)=-2(\ee^{\VddVT}-1)\sin\left(\frac{2\pi k}{N}\right)\,.
}
In the undriven (equilibrium) case $\Vdd=0$, all eigenvalues of $\mbb{M}$ are negative, implying that $\ve v^*$ is stable, as previously anticipated. However, as the voltage $\Vdd$ increases, there is a change in sign of $\mu_{k^*}$, i.e., the fixed point becomes unstable, for the pair of Fourier modes $\pm k^*$ (modulo $N$) with $k^*=(N-1)/2$ and $-k^*=(N+1)/2$. At the bifurcation where $\mu_{\pm k^*}=0$, the imaginary parts $\omega_{\pm k^*} = \pm\omega_{k^*}$ of the eigenvalues remain finite, which indicates a Hopf bifurcation~\cite{nicolis1995introduction,strogatz2024nonlinear}. The critical voltage $\VddcVT$ at which this bifurcation occurs can be expressed as
\begin{equation}\eqnlab{Vddstar}
    \frac{\Vdd^*}{V_\text{T}} = \log\left[1+\sec\left(\frac{\pi}{N}\right) \right]\,,
\end{equation}
for odd $N$, where $\sec(x) = 1/\cos(x)$. As $\sec(\pi/N)$ decreases with increasing number of inverters for $N\ge3$, the critical voltage $\Vdd^*$ decreases with increasing $N$.

If $\Vdd$ is increased even further, more pairs of Fourier modes $\pm k_m^*=[N\mp(2m-1)]/2$, where $m=2,\ldots,(N-1)/2$, become unstable through additional Hopf bifurcations. However, since these bifurcations emerge from the already unstable fixed point $\ve v^*$, they are much less relevant to the global dynamics than the primary bifurcation of the $k^*$ mode at $\Vddc$ in \Eqnref{Vddstar}, which renders the stable fixed point at $\ve v^*$ unstable. We therefore call $k^*$ the critical Fourier mode. Once this primary bifurcation has occurred, the system is driven away from $\ve v^*$ and towards other stable states in the long-time limit. The occurrence of stable states in the vicinity of the bifurcation is determined by the bifurcation's normal form, which we analyze in the following.
\subsection{Hopf normal form}\seclab{hnf}
In order to determine the behavior of the system close to the bifurcation at $\Vddc$, we consider small variations $\ve x = (\ve v - \ve v^*)/V_\text{T}$ around the fixed point $\ve v^*$ with $|\ve x|\ll1$. To this end, we Taylor-expand the deterministic dynamics \Eqnref{eom} up to $\mathcal{O}(\ve{x}^3)$, obtaining
\algn{\eqnlab{perturb_eom}
\tau_0 \dot x_i \sim &h_{10}x_i + h_{01}x_{i-1}+\frac16\left(h_{30}x^3_i+ 3h_{21} x_i^2x_{i-1}\right.\nn\\
		&\left.+3h_{12} x_ix^2_{i-1}+h_{03}x_{i-1}^3\right)\,,\nn\\
		= &-2[x_i + (\ee^{\VddVT}-1)x_{i-1}]+x_i^2x_{i-1}-x_ix_{i-1}^2\nn\\
		&-\frac13[x_i^3+(\ee^{\VddVT}-1)x_{i-1}^3]\,, 
}  
where we have neglected $\mathcal{O}(\ve x^4)$ terms and used $h_{20}=h_{11}=h_{02}=0$, $h_{30}=h_{12}=-h_{21}=-2$, and $h_{03} = -2(\ee^{\VddVT}-1)$. We observe that all quadratic terms in the expansion in \Eqnref{perturb_eom} vanish. This property of the model makes the calculations that follow significantly simpler because it will allow us to simply read off the Hopf normal form from the Fourier-transformed dynamics without explicit calculations.

To get there, we decompose the variations $\ve x$ into their discrete Fourier modes $\hat{\ve{x}} = (\hat x_0, \hat x_1, \ldots, \hat x_{N-1})^{\mathrm{T}}$ by $\ve{\hat x} = \mbb{F}\ve{x}$ using the matrix $\mbb{F}$ in \Eqnref{Ftrans}. We obtain
\algn{
	\tau_0 \dot{\hat x}_k &\sim(\mu_k +i \omega_k)\hat x_k +\frac1{N^2}\sum_{k',k''=0}^{N-1} \left[\frac{1}{6}(\mu_k+i\omega_k)\right.\nn\\
	&\left.+\ee^{\frac{i2\pi k'}{N}}\left(1-\ee^{\frac{i2\pi k''}{N}}\right) \right]{\hat x}_{k'}{\hat x}_{k''}{\hat x}_{k-k'-k''}\,. \eqnlab{fourier_eom}    
}
Since the variations $\ve x$ are real, the Fourier components $\hat x_k$ occur in complex conjugate pairs $z_k$ and $\Bar{z}_k$, such that
\begin{eqnarray}
z_k = \hat x_k ,\qquad \bar{z}_k = \hat x_{-k}\,.
\end{eqnarray}
As we have observed in \Secref{FP}, the trivial fixed point loses its stability when increasing $\Vdd$ across $\Vdd^*$ in \Eqnref{Vddstar}, showing characteristics of a Hopf bifurcation in the plane spanned by $\hat x_{k^*}$ and $\hat x_{-k^*}$. The parametric distance from the critical voltage $\Vdd^*$ is conveniently measured by the bifurcation parameter
\algn{
	\lambda =\frac{\Vdd - \Vddc}{V_\text{T}}\,.
}
For $\lambda>0$ close to the bifurcation, i.e., $\lambda\ll1$, the long-term dynamics of the system are predominantly governed by the growing critical Fourier mode $z_{k^*}$ with $k^*=(N-1)/2$, because $\mu_{k^*}>0$. All other modes  $z_{k\neq k^*}$, on the contrary, have $\mu_k<0$, and thus decay exponentially over time, leading to $z_k = \bar z_k = 0$ for $k\neq k^*$ in the long-time limit. Since these non-critical modes vanish, the dynamics near the bifurcation can be effectively described by the critical Fourier mode $k^*$ alone. 

\begin{figure}
	\includegraphics[width=\linewidth]{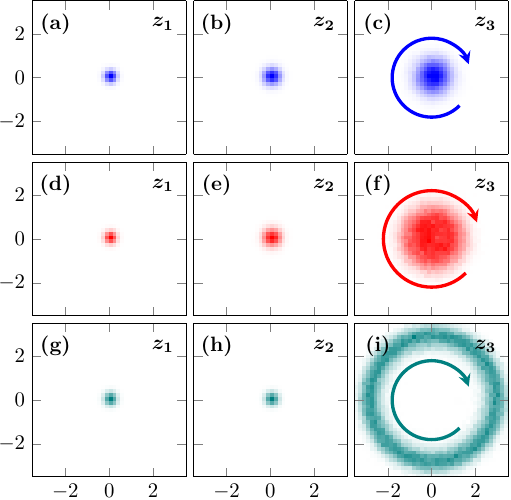}
	\caption{Steady-state probability density of Fourier modes $z_i$, $i=1,2,3$, for a $N=7$-stage ring oscillator below and above the critical voltage $\Vddc$, obtained from Gillespie simulations~\cite{gillespie1977exact} with $\Omega=10^2$, averaged over $10^2$ realizations and over a time interval $\Delta t\approx 100\tau_0$. The probability density is shown as a heat map in the complex plane, where the horizontal axes correspond to the real parts, the vertical axes to the imaginary parts of $z_i$. Darker colors indicate higher density, arrows indicate directions of rotation. \textbf{(a)}--\textbf{(c)} $\lambda=-0.1$. \textbf{(d)}--\textbf{(f)} $\lambda=0$. \textbf{(g)}--\textbf{(i)} $\lambda=0.2$.
}
	\label{fig:fourier_trajecs}
\end{figure}

Figure~\figref{fourier_trajecs} shows this behavior for $N=7$, so that $k^* = 3$: The figure shows the steady-state probability density of the Fourier modes $z_1$, $z_2$ and $z_3$ in the complex plane below and above the critical voltage $\Vddc$, i.e., for negative and positive $\lambda$, respectively. For $\lambda>0$ we clearly observe the emergence of the non-vanishing critical mode $z_k^*$ rotating in the $k^*=3$ plane, while the amplitudes of all other modes remain small.

In addition to the irrelevance of the non-critical modes, the analysis of $z_{k^*}$ is further simplified by studying the normal form of the Hopf bifurcation of $z_{k^*}$. This simplified form of \Eqnref{fourier_eom} represents an equivalent description of the dynamics close to the bifurcation but contains only the essential terms. In the general case, to bring \eqnref{perturb_eom} into normal form, one applies a non-linear coordinate transform that removes all non-essential terms from the cubic expansion in \Eqnref{fourier_eom}~\cite{guckenheimer1983nonlinear,golubitsky1988singularities}, as done explicitly in~\cite{meibohm2024minimum,meibohm2024small}. However, in the case at hand, this procedure is not necessary, because the quadratic terms in the expansion in \Eqnref{fourier_eom} vanish identically. This implies that coordinate transforms to remove the third-order inessential terms will produce only corrections of order four and higher~\cite{guckenheimer1983nonlinear,golubitsky1988singularities}, which we can safely neglect. Consequently, the Hopf normal form~\cite{nicolis1995introduction} for the bifurcation of $z_{k^*}$ can simply be read off \Eqnref{fourier_eom}, leading us to
\algn{\eqnlab{z_k_eom}
	\tau_0\,\dot{z}_{k^*} \sim \left(\ms{D}_{k^*}- \ms{C}_{k^*}|z_{k^*}|^2 \right) z_{k^*}\,.
}
The linear term $\ms{D}_{k^*} = \mu_{k^*} + i \omega_{k^*}$ in \Eqnref{z_k_eom} captures the instability of the fixed point $\ve v^*$, as discussed in \Secref{FP}. We expand $\ms{D}_{k^*}$ to linear order in $\lambda$, to obtain
\begin{align}
    \mu_{k^*} &= 4\cos^2\left(\frac{\pi}{2N}\right)\lambda + \mathcal{O}(\lambda^2)\,,  \label{eq:linear_real}\\
    \omega_{k^*} &= -\tan\left(\frac{\pi}{N}\right)\left(2 +  \mu_{k^*} \right) + \mathcal{O}(\lambda^2)\,. \label{eq:linear_im}
\end{align}
The properties of small-amplitude oscillations are encoded in the cubic term in \Eqnref{z_k_eom}. We divide $\ms{C}_{k^*}$ into real and imaginary parts, $\ms{C}_{k^*}=\ms{A}_{k^*}+i\ms{B}_{k^*}$. To leading order in $\lambda$, we find
\begin{align}
\ms{A}_{k^*} &= \frac{2}{N^2} \cos^2\left(\frac{\pi}{2 N}\right) \left[4\cos^2\left(\frac{\pi}{2N}\right)-1\right] + \mathcal{O}(\lambda) \label{eq:A_a}\\
\ms{B}_{k^*} &= -\frac{1}{N^2} \left[\cos\left(\frac{\pi}{N}\right) + \cos\left(\frac{2\pi}{N}\right)\right] \tan\left(\frac{\pi}{N}\right) + \mathcal{O}(\lambda)\,. \label{eq:B_a}
\end{align}
To decompose the Hopf normal form [Eq.~\eqnref{z_k_eom}] into radial and phase dynamics, we transform to polar coordinates $(r_{k^*},\varphi_{k^*})$, such that $z_{k^*} = r_k \ee^{i\varphi_{k^*}}$, leading us to:
\begin{align}
    \tau_0\dot{r}_{k^*} &\sim \bigg(\mu_{k^*} - \ms{A}_{k^*}r_{k^*}^2 \bigg) r_{k^*}\,, \label{eq:eom_radial}\\
    \tau_0\dot{\varphi}_{k^*} &\sim \omega_{k^*} - \ms{B}_{k^*} r_{k^*}^2\,. \label{eq:eom_phase}
\end{align}
The structure of radial dynamics~\eqnref{eom_radial} admits an additional attractor with $\dot{r}_{k^*} =0$, $\dot{\varphi}^*_{k^*}\neq0$ for some finite $r^*_{k^*}>0$, in addition to $r_{k^*} = 0$, which corresponds to the trivial fixed point $\ve v^*$. This additional attractor corresponds to a limit cycle, characterizing the oscillations observed in \Figsref{sample_trajecs} and \figref{fourier_trajecs}.
\begin{figure}
\centering
\includegraphics[width = \linewidth]{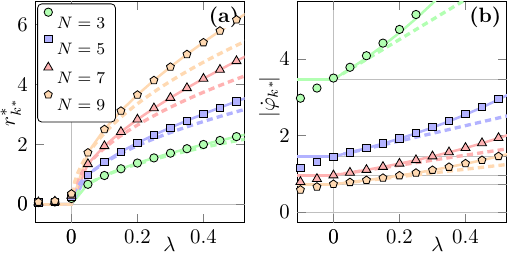}
\caption{Small-amplitude oscillations in the thermodynamic limit as function of $\lambda$ and for different numbers of inverters $N$. (a) Amplitude $r^*_{k^*}$ from \Eqnref{rdec} (dashed lines), from numerical solution of \Eqsref{eom} (solid lines), and from Gillespie simulations~\cite{gillespie1977exact} with $\Omega=10^4$ (symbols). (b) Angular velocity $|\varphi^*_{k^*}|$ from \Eqnref{mod_freq} (dashed lines), from numerical solution of \Eqsref{eom} (solid lines), and from Gillespie simulations with $\Omega=10^4$ (symbols).}
\label{fig:radius_det}
\end{figure}
To analyze this limit cycle, we first obtain for $\lambda>0$ the positive fixed point $r_{k^*}^*$ of Eq.~\eqref{eq:eom_radial} as,
\begin{equation}\eqnlab{rdec}
     r^{*2}_{k^*} \sim \frac{\mu_{k^*}}{\ms{A}_{k^*} } \sim \frac{N^2}{2\left[\cos^2\left(\frac{\pi}{2N}\right) - \frac14\right]}\lambda\,.
\end{equation}
Hence, we find that for $\lambda\ll1$, the amplitude of the oscillations increases with the applied voltage and the number of inverters as $r^{*}_{k^*} \propto N\sqrt{\lambda} = N\sqrt{(\Vdd- \Vdd^*)/V_\text{T}}$.

Figure~\ref{fig:radius_det}(a) shows $r^*_{k^*}$ as a function of $\lambda$ close to the bifurcation as obtained from \Eqnref{rdec}, compared with Gillespie simulations~\cite{gillespie1977exact} of the stochastic voltage vector $\ve v$ at large systems size and with numerical solutions of the deterministic dynamics~\eqnref{eom}. We observe good agreements between all methods for $\lambda\ll1$.

Similarly, the angular velocity $\dot \varphi_{k^*}$ experiences corrections of order $\lambda$ for $\lambda>0$, as the system is driven away from the unstable fixed point. These angular velocity corrections can be computed from \Eqnref{eom_phase}, we find
\algn{\eqnlab{mod_freq}
	\tau_0\dot \varphi_{k^*} \sim \omega_{k^*}|_{\lambda=0}\left[1+\frac{\cos^2\left(\frac{\pi}{2N}\right)}{\cos^2\left(\frac{\pi}{2N}\right) - \frac14}\lambda\right]\,.
}
The second term in the brackets is positive for all odd $N\geq3$, suggesting that the frequency of oscillations increases with increasing $\lambda$, to leading order in $\lambda$. This finding is confirmed in \Figref{radius_det}(b), which shows $|\dot \varphi_{k^*}|$ from Gillespie simulations, from a numerical solution of \Eqnref{eom}, and from the theory in \Eqnref{mod_freq}. The numerical results are in good agreement with the theory for small positive $\lambda$. For larger $\lambda$, the frequency-increasing effect of oscillations becomes even stronger due to higher-order corrections in $\lambda$, which are not captured by \Eqnref{mod_freq}.
\subsection{Stability}\seclab{stab}
The stability of oscillating states is determined by a linear stability analysis of Eq.~\eqref{eq:eom_radial}. The limit cycle corresponding to $r_{k^*}^*$ is stable if the derivative $\mbb{N}_{k^*}\equiv\partial \dot r_{k^*}/\partial r_{k^*}|_{r_{k^*}=r^*_{k^*}}$ of the amplitude flow~\eqnref{eom_radial} is negative when evaluated on the limit cycle, i.e.,
\begin{equation}
    \mbb{N}_{k^*} = (\mu_{k^*} - 3\ms{A}_{k^*}{r^*_{k^*}}^2) = -2\mu_{k^*}<0\,.\eqnlab{lc_stab}
\end{equation}
Equation~\eqnref{lc_stab} confirms that the limit cycle is stable with respect to transverse disturbances for $\lambda>0$. Along the direction along the limit cycle, the stability exponent vanishes, $\partial \dot \varphi_{k^*}/\partial \varphi_{k^*} = 0$, which means that longitudinal disturbances neither grow nor shrink. This enables free diffusion along the limit cycle, leading to long-time decoherence of the phase $\varphi_{k^*}$, which we explore in \Secref{phasedec}.

A simple measure for the overall local stability of a given state is the so-called phase-space contraction rate $\ms{L}$. It measures by how much infinitesimal phase-space volumes are contracted in the vicinity of a given state. We define the dimensionless phase-space contraction rate as~\cite{ott2002chaos}
\algn{
	\ms{L} = -\tau_0\sum_{n=0}^{N-1}\frac{\partial \dot{v}_n}{\partial v_n} \,.
}
Using the dynamics~\eqnref{perturb_eom} close to the bifurcation and the Fourier transform, we can write $\ms{L}$ as
\algn{\eqnlab{pscont}
	\ms{L}\sim \ms{L}_0 + \frac{8}{N}\cos^2\left(\frac{\pi}{2N}\right)r_{k^*}^{*2}\,,
}
where $\ms{L}_0 = 2N$ is the phase-space contraction rate at the fixed point. Equation~\eqnref{pscont} shows that the rate of phase-space contraction increases in the presence of small amplitude oscillations for $\lambda>0$. We define the relative change $\Delta\ms{L}$ of the phase-space contraction rate as
\algn{
	\Delta\ms{L} &= \frac{\ms{L}-\ms{L}_0}{|\ms{L}_0|} \sim \frac{4}{N^2}\cos^2\left(\frac{\pi}{2N}\right)r_{k^*}^{*2}\nn\,,\\
	&\sim  \frac{2\cos^2\left(\frac{\pi}{2N}\right)}{\cos^2\left(\frac{\pi}{2N}\right) - \frac14}\lambda>0\,,\eqnlab{DL}
}
for $\lambda>0$. We will return to this relation in \Secref{epr}.
\subsection{Entropy production rate}\seclab{epr}
Small-amplitude oscillations for $\lambda>0$ are facilitated by a constant non-equilibrium drive across each inverter. This drive keeps the system far from equilibrium and renders the dynamics irreversible. This irreversibility in turn is characterized by a finite rate of irreversible entropy production $\dot{\Sigma}$. In the large system-size limit ($\Omega\to \infty$), the time-averaged entropy production is equal to the time-averaged heat dissipated by the transistors~\cite{freitas2021stochastic,meibohm2024small}. Since the currents through the transistors and thus the entropy production scale with the size of the system $\Omega$, it is convenient to introduce the intensive entropy production rate $\dot{\sigma} \equiv \dot{\Sigma}/ \Omega $.

Following the previous discussion, $\dot\sigma$ can be directly computed from the macroscopic transition rates $\omega_{i\pm}^{\text{p/n}}$~\cite{falasco2025macroscopic}, averaged over a period $T$ of the limit cycle:
\algn{
	&\dot \sigma =\frac{\tau_0}T\sum_{i=0 }^{N-1}\sum_{\rho=\text{n},\text{p} }\int_0^T\!\!\!\ed t\left[\omega^{\rho}_{i+} (\ve v) - \omega^{\rho}_{i-}(\ve v)\right]\log\frac{\omega^{\rho}_{i+} (\ve v)}{\omega^{\rho}_{i-}(\ve v)}\,,\nn\\
    &= \left(\frac{2\Vdd}{TV_\text{T}}\right)\!\sum_{n=0}^{N-1}\!\int_0^T\!\!\!\ed t\left[\ee^{\VddVT}\!\cosh(v_{n-1})\! - \! \cosh(v_n\!-\!v_{n-1})\right]\,. \eqnlab{dsigma}
}
In this dimensionless formulation, the $\Omega$-scaled entropy production rate $\dot \sigma$ is measured in units of $\kB/\tau_0$, where $\kB$ denotes Boltzmann's constant. Close to the bifurcation, for $\lambda>0$, $\lambda\ll1$, we have
\algn{
	\dot \sigma 	&\sim \dot \sigma_0 + \left(\frac{2\Vdd}{V_\text{T}N}\right)\left[\ee^{\VddVT}-4\cos^2\left(\frac{\pi}{2N}\right)\right]r_{k^*}^{*2}\,,\nn\\
					&\sim \dot \sigma_0\left\{1+\left[\frac12 - \cos\left(\frac{\pi}{N}\right)\right]\frac{4}{N^2}\cos^2\left(\frac{\pi}{2N}\right)r_{k^*}^{*2}\right\}\,.\eqnlab{dsig}
}
where $\dot\sigma_0 = \left(2\Vdd N/V_\text{T}\right)\left(\ee^{\VddVT}-1\right)$ denotes the entropy production rate at the fixed point $\ve v^*$. Note that the time average in \Eqnref{dsigma} disappears close to the bifurcation [see \Eqnref{dsig}], because $r^*_{k^*}$ remains constant on the limit cycle to linear order in $\lambda$.

As $1/2 - \cos\left(\pi/N\right)<0$ for $N>3$, \Eqnref{dsig} shows that dissipation is reduced compared to $\dot \sigma_0$ by the presence of small-amplitude oscillations. For $N=3$, one has $1/2 - \cos\left(\pi/3\right) = 0$, so entropy production does not change to linear order in $\lambda$ close to the transition.

\begin{figure}
\includegraphics[width=\linewidth]{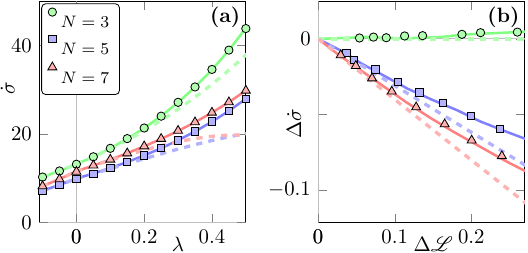}
\caption{(a) Entropy production rate $\dot \sigma$ as function of $\lambda$ for different numbers $N$ of inverters from \Eqnref{dsig} (dashed lines), from numerical solutions of \Eqnref{dsigma}, and from Gillespie simulations with $\Omega=10^3$ (symbols). (b) Stability-dissipation relation~\eqnref{sdr} compared with the numerical solution of \Eqsref{eom} (solid lines) and Gillespie simulations (symbols).}
\label{fig:sdr}
\end{figure}

Figure~\figref{sdr}(a) shows the entropy production rate $\dot \sigma$ as a function of $\lambda$ obtained from \eqnref{dsig} compared with our numerics for different $N$. We observe that $\dot \sigma$ generally increases with increasing $\lambda$. However, the difference $\dot \sigma-\dot \sigma_0$ is negative for $N>3$, as predicted by \Eqnref{dsig}. The case $N=3$, by contrast, is somewhat special in that the linear contribution of small-amplitude oscillations in \Eqnref{dsig} vanishes. Furthermore, the numerical simulations in \Figref{sdr}(a), show that the higher-order terms are \textit{positive}, which means that dissipation \textit{increases} due to oscillations when $N=3$, in contrast to the opposite behavior for $N>3$.

Defining the relative change $\Delta\dot\sigma = (\dot\sigma - \dot\sigma_0)/|\dot\sigma_0|$ of entropy production across the bifurcation, and using \Eqnref{DL}, we obtain the stability-dissipation relation
\algn{\eqnlab{sdr}
	\Delta\dot\sigma \sim \left[\frac12 - \cos\left(\frac{\pi}{N}\right)\right]\Delta\ms{L} \,,
}
for $0<\lambda\ll1$. A similar relation has previously been derived for driven Potts models in Refs.~\cite{meibohm2024minimum,meibohm2024small}. It shows that stability and dissipation are linearly related close to the bifurcation, and that decreasing dissipation corresponds to increasing local stability for $N>3$. The linear relation~\eqnref{sdr} is confirmed by our numerics~\footnote{For finite-size discrete systems, the analogue of $\mc{L}$ is the so-called inflow rate~\cite{baiesi2015inflow}, which converges to $\mc{L}$ in the thermodynamic limit~\cite{meibohm2024small}.} in \Figref{sdr}(b), which also shows that higher-order effects further away from the oscillation transition lead to a relative increase in $\Delta\dot\sigma$ compared to \Eqnref{sdr}.

To conclude this Section, we have characterized the small-amplitude oscillations close to the critical voltage $\VddcVT$ for any odd number of inverters. We find that the dynamics of ring oscillators in the long-time limit ($t/\tau_0 \to \infty$) relaxes into a stable oscillating state, characterized by a limit cycle with an increased phase-space contraction rate. In terms of the voltages $v_n$, the small-amplitude oscillations are given by
\begin{equation}
    v_n(t)= (-1)^n\frac{2 V_\text{T}}{N}r^*_{k^*}\cos\left(\dot\varphi_{k^*} t  - \frac{n\pi}N\right) + \mathcal{O}(\lambda)\,,
    \label{eq:voltage_limit_cycle}
\end{equation}
where the amplitude $r^*_{k^*}$ and the frequency $\dot\varphi_{k^*}$ are obtained in Eqs.~\eqref{eq:rdec} and \eqref{eq:mod_freq}, respectively. For $N>3$, the presence of oscillations decreases dissipation but increases stability, as summarized by a stability-dissipation relation~\eqnref{sdr}.
\section{Deviations from thermodynamic limit}\seclab{deviations}
In our analysis of the thermodynamic limit in \Secref{detlim} we have neglected finite-size voltage fluctuations. Such fluctuations are, however, present in any finite system, and are expected to be particularly relevant close to the oscillation transition. Going beyond the thermodynamic limit, we consider large but finite systems under the influence of weak finite-size fluctuations.

To this end, we perturbatively expand the Master equation~\eqnref{mastereqnnonint} to next-to-leading order in the system size $\Omega$ to capture the leading fluctuations. Although not generally suited to characterize global properties, such as fluctuation theorems, which depend on fluctuations of all magnitudes~\cite{gopal2022large}, such an expansion faithfully describes Gaussian fluctuations around the deterministic limit cycle discussed in \Secref{hnf}.

Expanding~\eqnref{mastereqnnonint} up to $\mathcal{O}(1/\Omega)$, we obtain the Fokker-Planck equation
\begin{equation} \eqnlab{FPeqn}
     \partial_t P(\ve v,t)\sim-\nabla_{\ve v}\cdot[\ve F(\ve v)P(\ve v,t)]+\frac{1}{\Omega}\nabla_{\ve v}^{\textsf{T}}\left[\mbb{\Hat{D}}(\ve v)\nabla_{\ve v}P(\ve v,t)\right]\,,
\end{equation}
with diffusion matrix $\hat D_{ij}$ given by
\begin{equation}
    \Hat{D}_{ij} = \frac{1}{\tau_0}\left[\cosh{\left(\frac{v_i-v_{i-1}}{V_\text{T}}\right)}+\ee^{\VddVT}\cosh{\left(\frac{v_{i-1}}{V_\text{T}}\right)}\right]\delta_{ij}\,,
\end{equation}
which is in diagonal form for the given model.  Equation~\eqnref{FPeqn} corresponds to the following stochastic differential equation for the voltage vector $\ve v$:
\begin{align}\eqnlab{Strat_Langevin_eqn}
 \frac{\ed v_i}{V_\text{T}} &= F_i(\ve v)\ed t + \sqrt{\frac{2 \hat D_{ii}(\ve v)}{\Omega}} \circ \ed W_i\,.
\end{align}
Here, $\circ \ed W_i$ are independent Wiener increments in the Stratonovich convention~\cite{gardiner1985handbook}, with $\langle \ed W^2_i(t)\rangle = \ed t$. 
\subsection{Stochastic normal form}
To analyze \Eqnref{Strat_Langevin_eqn} further, we bring it into the stochastic version of the normal form~\eqnref{z_k_eom}, sometimes called ``stochastic normal form''~\cite{xiao2007effects}, which captures the fluctuations to leading order, close to the deterministic normal-form behavior.

In the weak-noise limit close to the bifurcation, there are two small parameters: The distance $\lambda$ from the bifurcation and the inverse system size $\Omega^{-1}$, which fixes the noise strength. To identify the leading noise-induced effects close to the bifurcation, we expand $\hat D_{ij}$ to the lowest order in $\lambda$, so that $D_{ii}(\ve v) \sim (1+\ee^{\VddcVT})/\tau_0$. Taking the discrete Fourier transform of Eq.~\eqnref{Strat_Langevin_eqn} and using the normal form in~\Secref{hnf} for $\ve F$, we obtain the stochastic equation
\algn{\label{Langevin_zk_1}
\ed z_{k^*} &\sim \tau_0^{-1}\bigg(\ms{D}_{k^*} - \ms{C}_{k^*}|z_{k^*}|^2 \bigg) z_{k^*} \ed t + \sqrt{\frac{2\bar D}{\Omega}}\circ \ed \hat W_{k^*}\,,
}
where $\bar{D}=N(1+\ee^{\VddcVT})/(2\tau_0)$. Furthermore,

\algn{
	\ed\hat W_{k^*} = \sqrt{\frac2{N}}\sum_{n=0}^{N-1}\ee^{\frac{i2\pi n k^*}{N}}  \circ \ed W_n\,,
}
is the weighted sum of $N$ independent Wiener increments, which is itself a complex Wiener increment, $\ed \hat W_{k^*} = \ed\hat W^{\Re}_{k^*} + i \,\ed\hat W^{\Im}_{k^*}$, with $\langle \ed\hat W^{\Re}_{k^*}(t)^2\rangle = \langle \ed\hat W^{\Im}_{k^*}(t)^2\rangle = \ed t$ and $\langle \ed\hat W^{\Re}_{k^*}(t)\ed\hat W^{\Im}_{k^*}(t)\rangle = 0$.

We again transform into polar coordinates $(r_{k^*},\varphi_{k^*})$ to obtain the stochastic phase-amplitude equations in the presence of finite-size fluctuations. For computational convenience, we first transform the coordinates and then move from the Stratonovich to the It\^o convention~\cite{gardiner1985handbook} to simplify the manipulations that follow.

In this way, we obtain stochastic phase-amplitude equations, given by
\begin{subequations}\eqnlab{snf}
\algn{
\ed r_{k^*} &\sim \tau_0^{-1}\left(\mu_{k^*} r_{k^*} - \ms{A}_{k^*}r_{k^*}^3 +  \frac{\tau_0\bar{D}}{r_{k^*} \Omega} \right) \, \ed t +\sqrt{\frac{2\bar{D}}{\Omega}}\cdot \ed\hat W^r_t\label{eq:stochastic_radial},\\
\ed\varphi_{k^*} &\sim \tau_0^{-1}\left(\omega_{k^*} - \ms{B}_{k^*}r^2_{k^*}\right)\ed t +\frac{1}{r_{k^*}}\sqrt{\frac{2\bar{D}}{\Omega}}\cdot \ed\hat W^\varphi_t \label{eq:stochastic_phase}\,,
}
\end{subequations}
where $\cdot \ed\hat W^r_t$ and $\cdot \ed\hat W^\varphi_t$ are independent Wiener increments in It\^o convention with $\langle {\ed\hat W^r_t}^2 \rangle = \langle {\ed \hat W^\varphi_t}^2 \rangle = \ed t$ and $\langle \ed \hat W^r_t \ed\hat W^\varphi_t \rangle=0$, which characterize finite-size fluctuations.

The presence of finite-size noise in the phase-amplitude equations has important consequences that we discuss in the following.

\subsection{Noise-induced oscillations}\seclab{nio}
In the presence of finite-size voltage fluctuations, the system oscillates even for $\lambda\leq0$, i.e., below and at the bifurcation. Such noise-induced oscillations arise as a combination of finite-size noise and the system's oscillatory relaxation towards its fixed point. More precisely, for $\lambda\leq0$, finite-size voltage fluctuations drive the system away from the stable fixed point, by means of the noise-induced term $\bar{D}/(r_{k^*}\Omega)$ in \Eqnref{stochastic_radial}. This term, together with the noise term in \Eqnref{stochastic_radial}, vanishes as $\Omega \to \infty$, resulting in the deterministic dynamics discussed in \Secref{detlim}.

As a consequence of this, oscillations are no longer characterized by a single closed limit cycle as in \Secref{detlim} but by a probability density of possible amplitudes $r_{k^*}$ and phases $\varphi_{k^*}$. To quantify the amplitudes of noise-induced fluctuations in the long-time limit, we first compute the steady-state probability density $P_s(r_{k^*})$ from the Fokker-Planck equation corresponding to \Eqnref{stochastic_radial}. This leads us to
\algn{\eqnlab{Pss_r}
    P_s(r_{k^*}) = \ms{N}^{-1} r_{k^*}\, \exp\left[-\frac1{\zeta}\left(\frac{\ms{A}_{k^*}}{\mu_{k*}}r_{k^*}^2- 1 \right)^2\right]\,,
}
where the normalization $\ms{N}$ is given by
\algn{
	\ms{N} = \frac{\sqrt{\pi}}{4}\frac{\zeta|\mu_{k^*}|}{\ms{A}_{k^*}}\left[\text{erf}\left(\frac1{\zeta}\right)+1\right]\,.
}
In \Eqnref{Pss_r}, we have defined the dimensionless parameter
\begin{equation}
	 \zeta \equiv \sqrt{\frac{4\ms{A}_{k^*}\bar D\tau_0}{\Omega\mu_{k^*}^2}}\,,
\end{equation}
which quantifies the importance of finite-size fluctuations. Sufficiently far away from the bifurcation for $\mu_{k^*}\propto\lambda\gg\Omega^{-1/2}$, fluctuations are mostly irrelevant, and we have $\zeta\ll1$. In this case, \Eqnref{Pss_r} strongly focusses around the maximum of the distribution. Close to the bifurcation $\lambda\ll\Omega^{-1/2}$, by contrast, we have $\zeta\gg1$. In this case, finite-size fluctuations are important even as the system size tends to infinity.

In particular, the significance of $\zeta$ is understood by comparing the most likely amplitude $r^{ss}_{k^*}$ with the magnitude of fluctuations around $r^{ss}_{k^*}$. The most likely amplitude $r^{ss}_{k^*}$ is obtained from \Eqnref{Pss_r}, by solving $P'_s(r^{ss}_{k^*})=0$ for $r^{ss}_{k^*}$, which gives
\begin{equation}\eqnlab{r_most_prob}
    {r^{ss}_{k^*}}^2 =\frac{|\mu_{k^*}|}{2\ms{A}_{k^*}}\left[\text{sign}(\lambda) +\sqrt{1 + \zeta^2}\right]\,.
\end{equation}
In the deterministic limit $\zeta\to0$, \Eqnref{r_most_prob} reduces to ${r^{ss}_{k^*}}^2=0$ for $\lambda<0$ and ${r^{ss}_{k^*}}^2=\mu_{k^*}/\ms{A}_{k^*}$ for $\lambda>0$, in agreement with \Eqnref{rdec}. For $\zeta\gg1$, i.e., close to the bifurcation, we find ${r^{ss}_{k^*}}^2 \sim \sqrt{\bar D\tau_0/(\Omega \ms{A}_{k^*})}>0$. Hence, as advertised earlier, finite-size fluctuations may induce coherent oscillations of finite amplitude, close to and even at the bifurcation, which are absent in the deterministic dynamics in \Secref{detlim}.

The magnitude of Gaussian fluctuations $\sigma_{k^*}^{ss}$ around $r^{ss}_{k^*}$ is estimated from ${\sigma^{ss}_{k^*}}^{-2}\sim P''_s(r^{ss}_{k^*})$, which gives
\algn{\eqnlab{sigss}
	{\sigma^{ss}_{k^*}}^2 \sim \frac{|\mu_{k^*}|}{2\ms{A}_{k^*}}\frac{\zeta^2}{4(1+\zeta^2)^{\frac12}}\,.
}
This expression shows that in the deterministic limit $\zeta\to0$, fluctuations  around $r_{k^*}^{ss}$ (of order $\sigma^{ss}_{k^*}$), decay as $\zeta\propto\Omega^{-1/2}$. Close to the bifurcation, for $\zeta\gg1$, we have ${\sigma^{ss}_{k^*}}^2\sim \sqrt{\bar D\tau_0/(\Omega \ms{A}_{k^*})}/4$ which is of the same order as ${r^{ss}_{k*}}^2$. This implies that when $\zeta$ is large, fluctuations play a significant role, regardless of the system size, which is a common property of fluctuations close to a phase transition~\cite{goldenfeld2018lectures}.
\begin{figure}
\includegraphics[width=\linewidth]{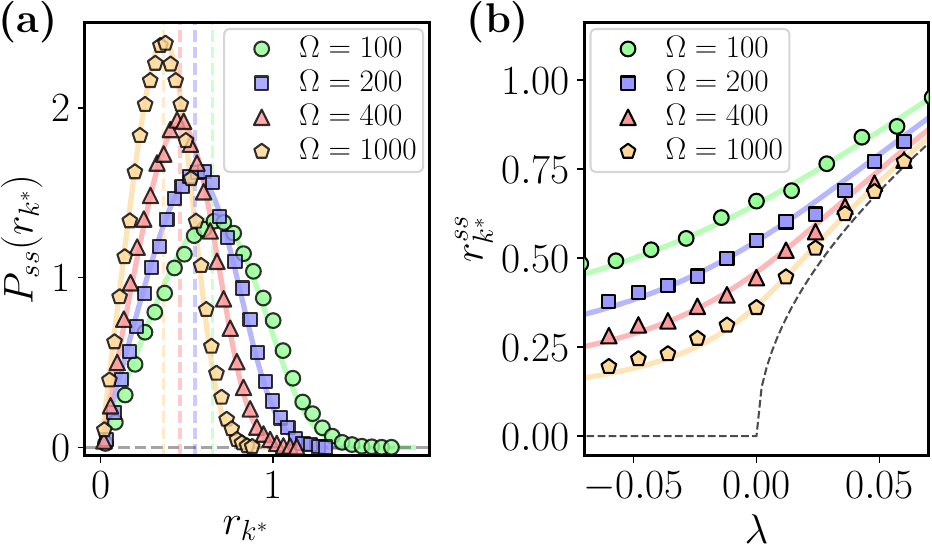}
\caption{Finite-size fluctuations close to the bifurcation for $N=3$ and different $\Omega$. (a) Steady-state probability density $P_s(r_{k^*})$ at $\lambda=0$ obtained from \Eqnref{Pss_r} (lines) and from Gillespie simulations (symbols). (b) Most likely radius $r_{\mathrm{ss}}^{\ast}$ in the steady state as a function of $\lambda$ from \Eqnref{r_most_prob} (lines) and Gillespie simulations (symbols). The dashed black line corresponds to the thermodynamic limit ($\Omega \to 
\infty$) from \Eqnref{rdec}. }
\label{fig:radial_dist}
\end{figure}

Figure~\ref{fig:radial_dist}(a) shows $P_s(r_{k^*})$ obtained using Eq.~\eqref{eq:Pss_r} and from Gillespie simulations at the bifurcation $\lambda=0$ ($\zeta\to\infty$), for $N=3$ and for different system sizes $\Omega$. We observe that \Eqnref{Pss_r} vanishes for small and large values of $r_{k^*}$ and exhibits a maximum around its most probable value, which decreases as the thermodynamic limit $\Omega\to\infty$ is approached. In all cases, \Eqnref{Pss_r} accurately captures the steady-state density. Along with the maximum value, the magnitudes of fluctuations around $r^{ss}_{k^*}$ decrease. As we have shown in \Eqsref{r_most_prob} and \eqnref{sigss}, the ratio between $r^{ss}_{k^*}$ and $\sigma^{ss}_{k^*}$ converges to a fixed value of order unity, $\sigma^{ss}_{k^*}/r^{ss}_{k^*}\to2$ as $\Omega\to\infty$, indicating that fluctuations remain important close to the bifurcation at arbitrary system size.

Figure~\ref{fig:radial_dist}(b) shows $r^{ss}_{k^*}$ as a function of $\lambda$ for $N=3$ and varying system size, from simulations and \Eqnref{r_most_prob}. In good agreement between theory and simulations, we observe that $r^{ss}_{k^*}$ approaches the deterministic amplitude $r^*_{k^*}$~\eqnref{rdec} as $\Omega$ increases at fixed $\lambda$, i.e., $\zeta$ decreases. Finally note that, for increasing $\Omega$, the region around the bifurcation where fluctuations are relevant, i.e., $\zeta$ is not small, shrinks as $\Omega^{-1/2}$.
\subsection{Phase decoherence}\seclab{phasedec}
As mentioned in the context of the stability analysis in \Secref{stab}, finite-size fluctuations of the phase $\varphi$ can propagate unobstructedly in the longitudinal direction along the limit cycle. This causes $\varphi_{k^*}$ to diffuse freely along the limit cycle, leading to a long-term decoherence of the phase. This phase decoherence is harnessed when ring oscillators are used as true random-number generators~\cite{robson2014truly,bucci2003high, sunar2006provably}, because the phase difference between several ring oscillators is sensitive to ``truely random'' finite-size fluctuations. In this context, the rate at which independent, true random numbers can be generated is related to the decoherence time of $\varphi_{k^*}$. We now estimate this decoherence time from the stochastic normal form~\eqnref{snf}.

The decoherence time $\tau_c$ of stochastic oscillations is quantified by the inverse exponential rate at which the auto-correlation function $\mathcal{C}_n(\tau)=\langle v_{n}(t) v_n(t+ \tau) \rangle$ decays as a function of $\tau$ for $\tau,t\gg\tau_0$. In other words, the decoherence time $\tau_c$ estimates the time over which an ensemble of initially coherent oscillators loses its coherence due to phase noise, eventually leading to uniform randomness of the output $v_n$.

At large but finite system size $\Omega\gg1$, we obtain an expression for $\tau_c$ by expanding the radial dynamics around the most probable value $r_{k^*}^{ss}$ of the stationary distribution. In Appendix~\ref{apsec:decoherence}, we show that such an expansion can be consistently formulated as long as the parameter
\algn{\eqnlab{bareps}
	\bar \eps \equiv f_{\text{sign}(\lambda)}\left(\zeta\right)\,,
}
becomes small for large system size, i.e., $\bar \eps \ll1$ for $\Omega\gg1$. The function $f_\pm$ has the form
\algn{
	f_\pm(\zeta) = \frac{\zeta}2\left(\sqrt{1+\zeta^2}\pm2\right)\left(\sqrt{1+\zeta^2} \pm 1\right)^{-\frac12}(1+\zeta^2)^{-\frac34}\,.
}
From an analysis of $f_\pm(\zeta)$ we find that $\bar \eps$ becomes small sufficiently far away from the bifurcation for $\zeta\ll1$. In this case, one has
\algn{
	\bar \eps = f_+(\zeta) \sim \frac{3\zeta}{\sqrt{8}}\ll1\,.
}
This finding is intuitive because $\zeta\ll1$ corresponds to the case when fluctuations are small, as discussed in \Secref{nio}.

In the opposite limit, for $\zeta\gg1$, i.e., close to the bifurcation, $\bar\eps$ remains of order unity even as $\Omega\to\infty$. The same holds outside the oscillating phase, for $\lambda<0$. In these regimes, expanding the dynamics around the maximum $r^{ss}_{k^*}$ is uncontrolled.

If $\bar \eps\ll1$, on the other hand, the dynamics can be systematically expanded around $r^{ss}_{k^*}$. Based on such an expansion, we show in Appendix~\ref{apsec:decoherence} that, to leading order, the decoherence time $\tau_c$ takes the form
\algn{\eqnlab{taudec}
	\tau^{-1}_c \sim \frac{\bar D}{\Omega}\left({r^{ss}_{k^*}}^{-2}+  \frac{\ms{B}_{k^*}^2 {r^{ss}_{k^*}}^{2}}{\ms{A}_{k^*}{r^{ss}_{k^*}}^{2} - \frac{\mu_{k^*}}{2}}\right)\,.
}
In the deterministic limit $\zeta\ll1$, this expression agrees with that found in~\cite{remlein2022coherence}, where the expansion was carried out around the deterministic radius $r^*_{k^*}$ in \Eqnref{rdec}, instead of $r^{ss}_{k^*}$. The first term in \Eqnref{taudec} corresponds to decoherence due to phase diffusion with diffusion constant $D_\varphi(r^{ss}_{k^*}) = \bar{D}/(\Omega {r^{ss}_{k^*}}^2)$ evaluated at the maximum $r^{ss}_{k^*}$ of the radial distribution. The second term is a consequence of weak interactions between the radial and phase dynamics~\cite{remlein2022coherence}.

In the oscillating state sufficiently far away from the bifurcation, when $\bar\eps\sim\zeta\ll1$, we have $r^{ss}_{k^*}\sim \lambda^{1/2}$, so that \Eqnref{taudec} predicts $\tau_c$ to grow linearly with system size, i.e. $\tau_c \sim \Omega$~\cite{gaspard2002correlation}. 

Close to the bifurcation for $\zeta\gg1$, \Eqnref{taudec} is not strictly valid, as it is based on an uncontrolled approximation. However, since the range of validity of \Eqnref{taudec} approaches the bifurcation at $\lambda=0$ as $\Omega\to\infty$, we speculate that \Eqnref{taudec} still predicts the dominant scaling of $\tau_c$ with $\Omega$, albeit possibly with wrong prefactor, in the limit of infinite system size. In the regime $\zeta\gg1$, $r_{k^*}^{ss}$ scales as $r_{k^*}^{ss} \sim \Omega^{-1/4}$, so that \Eqnref{taudec} predicts $\tau_c$ to increase as $\tau_c \sim \Omega^{1/2}$, substantially slower than away from the bifurcation. We confirm these scalings numerically below. Furthermore, these results are consistent with the findings in Ref.~\cite{nguyen2018phase}, where the authors numerically observed the same scalings close to and far away from bifurcations in several biochemical oscillator models.

\begin{figure*}
\centering
\includegraphics[width=\linewidth]{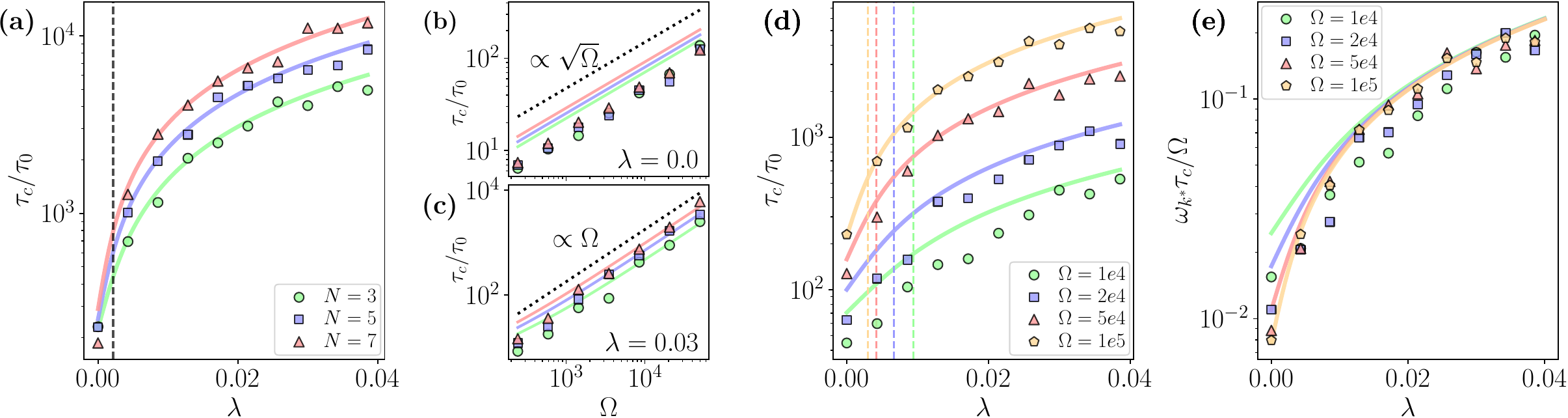}
\caption{Decoherence time $\tau_c$ as a function of different parameters. Symbols denote adaptive tau-leaping simulations~\cite{cao2007adaptive}. The solid lines are derived from \Eqnref{taudec} and \Eqnref{mod_freq}. The dashed vertical lines show where $\zeta\approx1$ for given $N$ and $\Omega$. (a) $\tau_c$ as a function of $\lambda$ for different number of inverters $N$ at system size $\Omega=10^5$. (b) $\tau_c$ as a function of the system size $\Omega$ at fixed $\lambda=0$, and (c) at fixed $\lambda=0.03$, for the same values of $N$. (d) $\tau_c$ as a function of $\lambda$ for different system sizes $\Omega$ with $N=3$ inverters. (e) Scaled quality factor $\omega_{k^*}\tau_c/\Omega$ as function of $\lambda$ for different system sizes $\Omega$ with $N=3$ inverters.}
\label{fig:decay_time_1}
\end{figure*}

Figure~\ref{fig:decay_time_1}(a) shows $\tau_c$ at finite but large system size $\Omega$ as function of $\lambda$ from \Eqnref{taudec} and from stochastic simulations with different numbers of inverters $N$. We observe good agreement between simulations and theory sufficiently far away from the bifurcation, i.e., to the right of the dashed line where $\zeta\lessapprox1$. The decoherence time $\tau_c$ increases with the number $N$ of inverters for fixed $\lambda$, due to the increase of the most-likely amplitude $r_{k^*}^{ss}$ in \Eqnref{taudec}. 

Figure~\ref{fig:decay_time_1}(b) shows the scaling of $\tau_c$ as a function of $\Omega$ for $\zeta\gg1$. The numerical data confirm our previous expectation that \Eqnref{taudec} should correctly predict the $\sim\Omega^{1/2}$ scaling close to the bifurcation, although the expansion around $r^{ss}_{k^*}$ is not controlled [$\bar \eps$ in \Eqnref{bareps} is not small]. However, the prefactor appears to be incorrect. Figure~\figref{decay_time_1}(c) shows the scaling away from the bifurcation, where $\zeta$ becomes small for the largest system sizes. In this case, both the $\sim\Omega$ scaling and the prefactor are correctly predicted by \Eqnref{taudec}.

This behavior is summarized for $N=3$ in \Figref{decay_time_1}(d), which shows $\tau_c$ as a function of $\lambda$ for a range of different system sizes. The vertical dashed lines show where $\zeta=1$. The figure shows how \Eqnref{taudec} becomes increasingly accurate with increasing system size, even close to $\lambda=0$.

Finally, we briefly comment on the relevance of the scaling properties of $\tau_c$ for the $\Omega$-scaled quality factor $\omega_{k^*}\tau_c/\Omega$~\cite{barato2016cost,nguyen2018phase,remlein2022coherence,santolin2025dissipation}, which is a measure for the number of coherent oscillations (scaled by $\Omega$) that the system is able to perform before losing its coherence. The $\Omega$-scaled quality factor has been shown to be bounded from above for thermodynamically consistent models by a constant of order unity~\cite{barato2016cost,remlein2022coherence,santolin2025dissipation}. Since $\omega_{k^*}$ becomes independent of $\Omega$ for large system size, see \Eqnref{mod_freq}, the $\Omega$-scaling of the quality factor $\omega_{k^*}\tau_c$ corresponds to the scaling of $\tau_c$. That is, after scaling with $\Omega$, $\omega_{k^*}\tau_c/\Omega$ tends to a constant for $\zeta\ll1$, away from the bifurcation. Close to the bifurcation for $\zeta\gg1$, by contrast, it approaches zero as $\Omega^{-1/2}$ for a large system size.

This behavior is shown in \Figref{decay_time_1}(e), which depicts $\omega_{k^*}\tau_c/\Omega$ as a function of $\lambda$ for different system sizes from theory and numerics. We observe that while the scaled quality factor becomes independent of the system size far from the bifurcation, it decreases with increasing system size in the vicinity of $\lambda=0$. In the entire $\lambda$ range that we considered, the quality factor is significantly smaller than unity, so thermodynamic bounds such as those derived in Refs.~\cite{barato2016cost,remlein2022coherence,santolin2025dissipation} are irrelevant here. This statement is true in particular close to the bifurcation, where oscillations are exceptionally decoherent, leading to a scaled quality factor that shrinks as $\Omega^{-1/2}$.
\section{Conclusions}\seclab{conc}
We analyzed small-amplitude oscillations and their fluctuations in $N$-stage CMOS ring oscillators operating in the subthreshold regime. Starting with a thermodynamically consistent model, we demonstrated that, in the thermodynamic limit $\Omega\to\infty$, oscillations emerge through a Hopf bifurcation resulting in a stable limit cycle when the supply voltage $\Vdd$ exceeds a critical threshold $\Vdd^*$. Exploiting the symmetries of the model, we used a normal-form analysis to obtain explicit expressions for the limit cycle and its frequency for any odd $N$, which are valid close to the bifurcation. Employing the framework of stochastic thermodynamics, we showed that the entropy production $\dot\sigma$ of ring oscillators decreases in the presence of oscillations for $N>3$, but increases sub-linearly for $N=3$. The change $\Delta\dot\sigma$ in the entropy production across the bifurcation is linearly related to the change in phase-space contraction $\Delta\ms{L}$ through a stability-dissipation relation [\Eqnref{sdr}], similar to that found in Refs.~\cite{meibohm2024minimum,meibohm2024small}.

For large but finite devices $\Omega = V_\text{T}/v_e \gg 1$, we characterized finite-size fluctuations using a systematic system-size expansion to capture the lowest-order effects of fluctuations around the deterministic limit cycle. Using a stochastic version of the Hopf normal form, we showed that finite-size fluctuations remain relevant close to the bifurcation, even in the thermodynamic limit, which is a common feature in the vicinity of phase transitions. In this regime, we showed that fluctuations induce finite-amplitude oscillations, even at and below the critical voltage. These noise-induced oscillations are characterized by an anomalously short decoherence time that was shown to scale sub-linearly $\tau_c\sim\Omega^{1/2}$ with $\Omega$.

This exceptional fast decoherence offers a route toward tunable randomness in probabilistic hardware, such as deterministic ``Ising machines''~\cite{sahoo2017ring,rahaman2025frequency,gonul2024multi} or stochastic probabilistic-bit systems~\cite{aadit2022physics}. Assessing the practical feasibility of such approaches and benchmarking them against existing subthreshold-CMOS p-bit implementations~\cite{jelinvcivc2025efficient}, remains an important direction for future work.

A key limitation of our analysis is that supply voltages must be small, of the order of the thermal voltage ($\sim 26\,\text{m}V$ at room temperature). Further investigations are required to determine whether the permutation symmetry used here allows for simplifications, even further away from the critical voltage.
\begin{acknowledgments}
This research was supported by the Project No. INTER/FNRS/20/15074473 funded by F.R.S.-FNRS (Belgium) and FNR (Luxembourg).
\end{acknowledgments}
\appendix
\section{Decoherence timescale calculation - Stochastic Normal Form}
\label{apsec:decoherence}
We present the details for the computation of the decorrelation time for the stochastic normal form~\eqnref{snf}. We again consider
\begin{subequations}\eqnlab{snf_stuart_landau}
\algn{
\ed r_{k^*} &\sim -\tau_0^{-1}V'(r_{k^*}) \, \ed t +\sqrt{\frac{2\bar{D}}{\Omega}}\cdot \ed\hat W_t^r\, \label{eq:ap_snf_sl_radius},\\
\ed \varphi_{k^*} &\sim \tau_0^{-1}\left(\omega_{k^*} - \ms{B}_{k^*}r^2_{k^*}\right)\ed t +\frac{1}{r}\sqrt{\frac{2\bar{D}}{\Omega}}\cdot \ed\hat W_{t}^\varphi \label{eq:ap_stochastic_phase}\,,
}
\end{subequations}
where
\algn{
	V(r) = -\frac{\mu_{k^*} r^2}2 + \ms{A}_{k^*}\frac{r^4}4 - \frac{\bar{D}\tau_0}{\Omega}\log(r)\,.
}
We proceed by expanding $V(r)$ around its minimum at $r^{ss}_{k^*}$, see \Eqnref{r_most_prob}. This yields for $V'(r_{k^*})$,
\begin{multline}
	V'(r_{k^*}) \sim V'(r^*) + V''(r^{ss}_{k^*})(r_{k^*}-r^{ss}_{k^*})\\
	 + \frac12 V'''(r^*)(r_{k^*}-r^{ss}_{k^*})^2\,.
\end{multline}
The minimum condition requires $V'(r^{ss}_{k^*})=0$, and we have
\algn{
	\kappa &\equiv V''(r^{ss}_{k^*}) =  2|\mu_{k^*}|\sqrt{1 + \zeta^2}\,,\\
	\xi &\equiv  V'''(r^{ss}_{k^*}) =  6\ms{A}_{k^*} r^{ss}_{k^*} -\frac{2\bar D\tau_0}{{r^{ss}_{k^*}}^3\Omega}\,,
}
We thus obtain the following expression for the dynamics of $\rho = r_{k^*}-r^{ss}_{k^*}$:
\algn{
	\ed\rho &\sim -\tau_0^{-1}(\kappa \rho + \xi \rho^2 )\, \ed t +\sqrt{\frac{2\bar{D}\tau_0}{\Omega}}\cdot \ed\hat W^r_t\, \label{eq:ap_snf_sl_radius_2}\,.
}
We rescale $\rho$ according to $\rho = \sqrt{\bar D\tau_0 /(\Omega\kappa)}\bar\rho$ and time as $t = \tau_0\bar t/\kappa$. This way, \Eqnref{ap_snf_sl_radius_2} depends on a single parameter. We obtain
\algn{\eqnlab{dbrho}
	\ed\bar\rho &\sim \left(-\bar\rho - \bar\eps \rho^2\, \right)\ed\bar t +\sqrt{2}\cdot \ed\hat W^r_{\bar t}\,\,,
}
with the parameter
\algn{
	\bar \eps \equiv\sqrt{\frac{\bar D\tau_0}{\Omega\kappa^3}}\xi\,,
}
which agrees with \Eqnref{bareps} when expressed in terms of $\zeta$. For small $\bar \eps$, we expand
\algn{\eqnlab{brhoexp}
	\bar\rho \sim \bar \rho^{(0)} + \bar\eps\bar \rho^{(1)} + \bar\eps^2\bar \rho^{(2)}+\ldots \,,
}
and evaluate the resulting equation~\eqnref{dbrho} order by order in $\bar\eps$. To linear order in $\bar \eps$, we find
\algn{
	\ed\bar\rho^{(0)} &\sim -\bar\rho^{(0)}\, \ed\bar t +\sqrt{2}\cdot \ed\hat W^r_{\bar t}\,\,,\\
	\ed\bar\rho^{(1)} &\sim -(1+2\bar\rho^{(0)})\bar\rho^{(1)}\, \ed\bar t\,\,.
}
To lowest order in $\bar \eps$, $\bar\rho^{(0)}$ is an standard Ornstein-Uhlenbeck process, while $\bar\rho^{(1)}$ is a non-Gaussian correction, that we disregard in the following. We note, however, that systematic expansions around the lowest order process $\bar \rho^{(0)}$ can be constructed in this way.

We now consider the correlation function in the long-time limit
\begin{align}
    C_n(\tau) = \langle \nu_n(\tau) \nu_n(0)\rangle \sim\left(\frac{V_\text{T}}{N}\right)^2 \mathcal{C}(\tau) + \text{c.c.}\,,
\end{align}
where $\mathcal{C}(\tau)= \langle z_{k^*}(\tau)\bar{z}_{k^*}(0)\rangle$ and c.c. denotes the complex conjugate. To study the exponential decay of $C_n(\tau)$ we thus evaluate 
\begin{align}
    \mathcal{C}(\tau) &= \langle r_{k^*}(\tau)r_{k^*}(0)\,\ee^{i[\varphi_{k^*}(\tau)-\varphi_{k^*}(0)]}\rangle\,.
\end{align}
Integrating \eqnref{ap_stochastic_phase}, we obtain
\begin{multline}
	\varphi_{k^*}(\tau)-\varphi_{k^*}(0) = \kappa^{-1}\!\!\int_0^{\tau \kappa}\!\!\!f[r_{k^*}(\bar t)]\ed \bar t\\ + \sqrt{\frac{2\bar\eps^2\kappa^2}{\xi^2}}\int_0^{\tau \kappa}\!\!\! g[r_{k^*}(\bar t)] \cdot \ed \hat W_{\bar t}^\varphi\,,
\end{multline}
where $f(r_{k^*}) = \omega_{k^*} - \ms{B}_{k^*}{r_{k^*}}^2$ and $g(r_{k^*}) = 1/r_{k^*}$.
In order to evaluate $\mathcal{C}(\tau)$, we first average over $\hat W_t^\varphi$, conditional on a fixed
$\{r_{k^*}(t)\}_{t\in[0,\tau]}$, leading to
\begin{widetext}
\begin{equation}
    \mathcal{C}(\tau) 
    = \Big\langle r_{k^*}(\tau)r_{k^*}(0)\Big\langle \ee^{i(\varphi_{\tau}-\varphi_0)} \,\Big|\, \{r_{k^*}(t)\} \Big\rangle_{\{\hat W^\varphi_{\bar t}\}}
        \Big\rangle_{\{r_{k^*}(t)\}} = \Big\langle r_{k^*}(\tau)r_{k^*}(0)
        \ee^{i \kappa^{-1}\!\!\int_0^{\tau\kappa} \!\! f[r_{k^*}(\bar t)]\ed \bar t - \frac{\bar\eps^2\kappa^2}{\xi^2}\int_0^{\tau\kappa}\!\! g^2[r_{k^*}(\bar t)]\ed \bar t }
                \Big\rangle_{\{r_{k^*}(t)\}} \,.
\end{equation}
\end{widetext}
We express $r_{k^*}(\bar t)$ as $r_{k^*}(\bar t) =  r^{ss}_{k^*} + \bar \eps (\kappa/\xi) \bar \rho(\bar t) $ and use the small-$\bar \eps$ expansion~\eqnref{brhoexp} to find to lowest non-vanishing order in $\bar\eps$
\algn{
 \mathcal{C}(\tau) \sim&  {r^{ss}_{k^*}}^2\ee^{\left[i\omega^{ss}_{k^*} - \frac{\bar\eps^2\kappa^3}{\xi^2}g^2(r^{ss}_{k^*})\right]\tau}\Big\langle
        \ee^{i \frac{\bar \eps}{\xi}f'(r^{ss}_{k^*})\!\!\int_0^{\tau\kappa} \!\! \bar \rho^{(0)}(\bar t)\ed \bar t}
                \Big\rangle_{\{\rho(\bar t)\}}\,,\\
\sim&{r^{ss}_{k^*}}^2\ee^{\left[i\omega^{ss}_{k^*} - \frac{\bar\eps^2\kappa^3}{\xi^2}g^2(r^{ss}_{k^*})\right] \tau -  \frac{\bar \eps^2}{2\xi^2}{f'(r^{ss}_{k^*})}^2 \text{Var}\left(\int_0^{\tau\kappa} \!\! \bar \rho^{(0)}(\bar t)\ed \bar t\right)}\,,
}
where $\omega^{ss}_{k^*} = f(r^{ss}_{k^*}) = \omega_{k^*} - \ms{B}_{k^*} {r^{ss}_{k^*}}^2$ corresponds to the oscillation frequency at $r^{ss}_{k^*}$. Evaluating $\text{Var}\left(\int_0^{\tau\kappa} \!\! \bar \rho^{(0)}(\bar t)\ed \bar t\right)$ for long times $\tau\kappa\gg1$ gives
\algn{
	\text{Var}\left(\int_0^{\tau\kappa} \!\! \bar \rho^{(0)}(\bar t)\ed \bar t\right) \sim 2\tau\kappa\,.
}
We thus obtain
\algn{
	\mathcal{C}(\tau) \sim {r^{ss}_{k^*}}^2\ee^{i\omega^{ss}_{k^*}\tau - \tau/\tau_c}\,,
}
where we have identified the decorrelation time $\tau_c$ as
\algn{
	\tau^{-1}_c \sim \frac{\bar\eps^2\kappa}{\tau_0\xi^2}\left[\kappa^2 g^2(r^{ss}_{k^*})+  {f'(r^{ss}_{k^*})}^2\right]\,,
}
to leading order in $\bar \eps$. Reinstalling the original parameters and substituting the expressions for $f$ and $g$ we thus obtain
\algn{
	\tau^{-1}_c \sim \frac{\bar D}{\Omega \kappa^2}\left[\frac{\kappa^2}{{r^{ss}_{k^*}}^2}+  4\ms{B}_{k^*}^2 {r^{ss}_{k^*}}^2\right]\,,
}
which agrees with \Eqnref{taudec} in the main text.
\end{document}